%% file: Z-MPI.tex
\documentclass[twocolumn,preprintnumbers,superscriptaddress,aps,nofootinbib]{revtex4-2}
\usepackage{amsmath,amssymb,mathtools}
\usepackage{graphicx}
\usepackage{booktabs,cellspace}
\usepackage{color}
\usepackage{hyperref}
\usepackage{xspace}
\usepackage[normalem]{ulem}

\usepackage{blindtext}

\newcommand{\mb}{\ensuremath{\,\mathrm{mb}}\xspace}
\newcommand{\fb}{\ensuremath{\,\mathrm{fb}}\xspace}
\newcommand{\pb}{\ensuremath{\,\mathrm{pb}}\xspace}

\newcommand{\GeV}{\ensuremath{\,\mathrm{GeV}}\xspace}
\newcommand{\TeV}{\ensuremath{\,\mathrm{TeV}}\xspace}

\newcommand{\sigmaeff}{\sigma_\text{eff}}

\newcommand{\logbook}[2]{}

\definecolor{darkgreen}{rgb}{0,0.4,0}
\definecolor{grey}{rgb}{0.5,0.5,0.5}
\definecolor{orange}{rgb}{0.9,0.5,0.0}
\usepackage[dvipsnames]{xcolor}

\newcommand{\OXaff}{Rudolf Peierls Centre for Theoretical Physics,
  Clarendon Laboratory, Parks Road, Oxford OX1 3PU, UK} 
\newcommand{\CERNaff}{CERN, Theoretical Physics Department, CH-1211 Geneva 23, Switzerland}
\newcommand{\ASCaff}{All Souls College, Oxford OX1 4AL, UK}
\newcommand{\DURaff}{Institute for Particle Physics Phenomenology, University
  of Durham, South Road, Durham DH1 3LE, UK}
\newcommand{\UZHaff}{Physik Institut, Universit\"at Z\"urich, CH-8057 Z\"urich, Switzerland}

\makeatletter 
    
\renewcommand\onecolumngrid{
\do@columngrid{one}{\@ne}%
\def\set@footnotewidth{\onecolumngrid}
\def\footnoterule{\kern-6pt\hrule width 1.5in\kern6pt}%
}

\renewcommand\twocolumngrid{
        \def\footnoterule{
        \dimen@\skip\footins\divide\dimen@\thr@@
        \kern-\dimen@\hrule width.5in\kern\dimen@}
        \do@columngrid{mlt}{\tw@}
}%

\makeatother    

\begin{document}

\title{Exploring high-purity multi-parton scattering at hadron colliders}

\preprint{CERN-TH-2023-055, DCPT/23/54, IPPP/23/27, OUTP-23-04P, ZU-TH 17/23}

\author{Jeppe R.~Andersen}       \affiliation{\DURaff}%
\author{Pier Francesco Monni}    \affiliation{\CERNaff}%
\author{Luca Rottoli}            \affiliation{\UZHaff}%
\author{Gavin P.\ Salam}         \affiliation{\OXaff}\affiliation{\ASCaff}%
\author{Alba Soto-Ontoso}        \affiliation{\CERNaff}  %

\begin{abstract}
  Multi-parton interactions are a fascinating phenomenon that occur in
  almost every high-energy hadron--hadron collision, yet are
  remarkably difficult to study quantitatively.
  In this letter we present a strategy to optimally disentangle
  multi-parton interactions from the primary scattering in a
  collision.
  That strategy enables probes of
  multi-parton interactions that are significantly beyond the state of the art,
  including their characteristic momentum scale, the interconnection
  between primary and secondary scatters, and the pattern of three and
  potentially even more simultaneous hard scatterings.
  This opens a path to powerful new constraints on multi-parton
  interactions for LHC phenomenology and to the investigation of their
  rich field-theoretical structure.
\end{abstract}

\maketitle

\begin{figure}
  \centering
  \includegraphics[width=0.6\columnwidth]{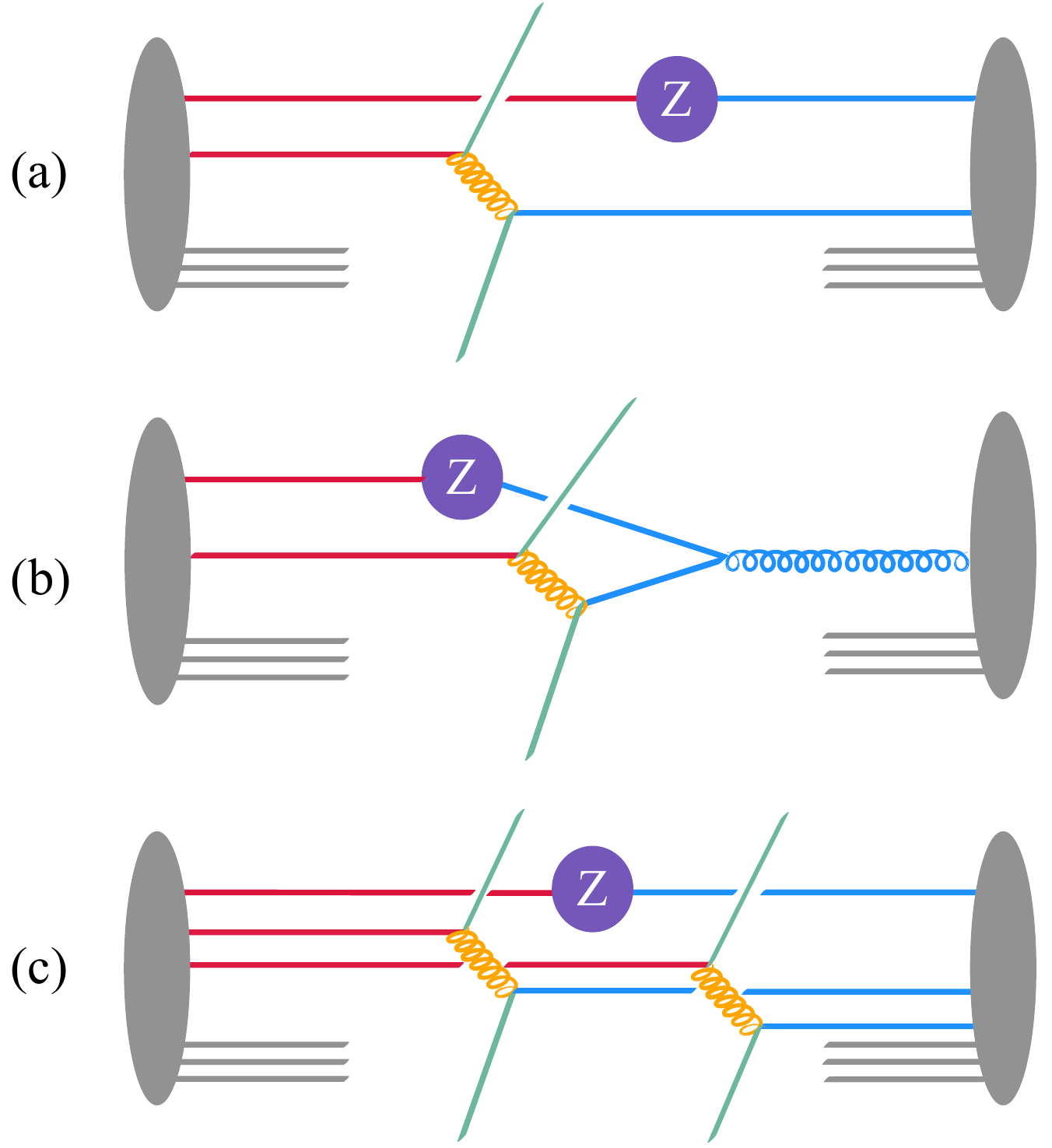}
  \caption{Illustration of some MPI configurations that will be probed
    in this paper:
    (a) standard double hard scattering, producing a
    $Z$ boson and a pair of jets;
    (b) perturbative interconnection between the partons involved in
    the two hard scatterings, where the $\bar q$ that produces
    the $Z$ and the $q$ that scatters to produce the dijet system both
    have a common origin in the perturbative splitting of a gluon;
    and (c) a process with three hard scatterings.
    %
  }
  \label{fig:mixing-hard-proc}
\end{figure}

At high-energy hadron colliders such as CERN's Large Hadron Collider
(LHC), almost every event that gets studied is accompanied by multiple
additional parton interactions (MPI) from the same proton--proton
collision, cf.\ Fig.~\ref{fig:mixing-hard-proc}. 
For example in each proton--proton collision that produces a $Z$ or Higgs
boson (the ``primary'' process),
models~\cite{Sjostrand:1987su,Sjostrand:2004pf} suggest that there are
about ten additional parton 
collisions that occur simultaneously, usually involving QCD
scatterings of quarks and gluons.
MPI are the subject of a rich array of studies, involving effects
ranging from partonic correlations inside the proton to colour
reconnections between final state quarks and
gluons~\cite{Jung:2016ymz,Bartalini:2018qje}.
Their modelling is an essential component of every major simulation
tool~\cite{Bierlich:2022pfr,Bellm:2019zci,Sherpa:2019gpd}.
While often thought of as non perturbative, we shall see clearly below
that MPI involve transverse momenta of up to $10\GeV$ and beyond, i.e.\ close to
the scale of many of the primary processes regularly studied at the
LHC.
%
%
In a context where there is an ambitious worldwide effort to bring
high precision in perturbative QCD calculations for those primary
process~\cite{Huss:2022ful,Caola:2022ayt,Campbell:2022qmc}, our
current partial understanding of MPI scatters risks becoming a
limiting factor across much of the LHC programme.

A major challenge in the experimental characterisation of MPI is the
difficulty of unambiguously separating the MPI signal from the primary
hard scattering.
In this letter we propose an approach to investigating MPI scatters in
Drell-Yan events that optimally suppresses the contamination from the
primary hard scattering.
This opens the path to a programme of experimental study of
MPI that
goes significantly beyond the current state of the art.
Features that we will highlight include (a) the clarity of the MPI
signal;
(b) scope for quantitative investigations of the leading two hard
scatters (2HS) that includes direct experimental sensitivity
to perturbative
interconnection~\cite{Diehl:2011tt,Blok:2011bu,Diehl:2011yj,Blok:2013bpa,Diehl:2017kgu}
between the primary process and the 
second hard scatter, cf.\ Fig.~\ref{fig:mixing-hard-proc}b;
and (c) the potential for  observation of high-purity triple
parton scattering
(Fig.~\ref{fig:mixing-hard-proc}c)~\cite{dEnterria:2016ids,CMS:2021qsn},
as well as sensitivity to even more than three scatters.

\begin{figure}
  \centering
  \includegraphics[width=\columnwidth,page=2]{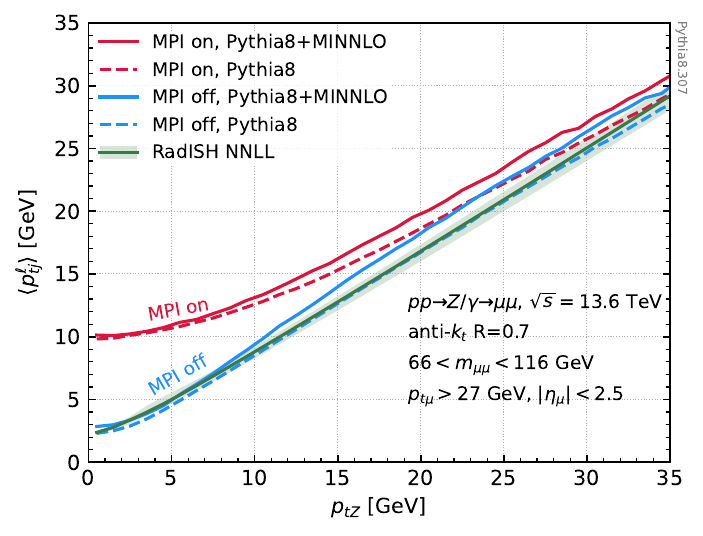}
  \caption{The average leading jet transverse momentum
    $\langle p_{tj}^\ell\rangle$ as a function of the $Z$ transverse
    momentum, in $Z \to \mu^+\mu^-$ events, with muon selection
    cuts as indicated in the plot.
    A radius of $R=0.7$ is used here to reduce the loss of transverse
    momentum from the jet due to final-state radiation and
    hadronisation and so more accurately track the transverse momentum
    of the underlying parton.
  }
  \label{fig:avg-jet-pt}
\end{figure}

The foundation of our approach is
the well-known
fact~\cite{Parisi:1979se} that if one considers events where the
Drell-Yan pair has a low transverse momentum, the amount of
initial-state radiation (ISR) is strongly constrained. 
To illustrate this quantitatively, we examine the average
transverse momentum of the leading jet in Drell-Yan events as a
function of the $Z$ transverse momentum.
Specifically, we consider $Z \to \mu^+ \mu^-$ events and cluster all
particles other than the muons with a jet algorithm (the anti-$k_t$
algorithm~\cite{Cacciari:2008gp} with a jet radius of $R=0.7$, as
implemented in FastJet~\cite{Cacciari:2011ma}).
Experimentally, this observable could be studied using charged-track
jets (see below), or possibly standard jets in a dedicated low-pileup run.
For now, to help expose the basic physical dynamics and scales, we
retain all particles in the jet clustering.
Fig.~\ref{fig:avg-jet-pt} shows results both without and with MPI.

Let us first concentrate on the curves without MPI: one is from a
resummed calculation (RadISH
NNLL~\cite{Monni:2016ktx,Bizon:2017rah,Monni:2019yyr}), the other from
a Monte Carlo simulation that uses a combination of
MiNNLO~\cite{Monni:2019whf,Monni:2020nks} with
POWHEG~\cite{Nason:2004rx,Frixione:2007vw,Alioli:2010xd} and
Pythia~8.3~\cite{Bierlich:2022pfr} (with HepMC2~\cite{Dobbs:2001ck}),
and the third is from Pythia alone.
All Pythia results use the Monash tune~\cite{Skands:2014pea}.
All three curves in Fig.~\ref{fig:avg-jet-pt} show the same features, namely that for almost the whole
range of $p_{tZ}$, the average leading jet $p_{t}$ is roughly
proportional to $p_{tZ}$ (with a proportionality coefficient close
to $1$), a consequence of momentum conservation between the jet and
the $Z$ boson.
For $p_{tZ}$ below about $2{-}3\GeV$ the average leading jet $p_t$
saturates.
Events with very small $p_{tZ}$ mostly occur when the
transverse recoil from one initial-state radiated gluon cancels with
that from other initial-state radiation.
In this region, the average leading jet $p_t$ has the parametric form
(\cite{supplement}, \S~\ref{app:scaling})
\begin{equation}\label{eq:ptj-lead}
\langle p_{tj}^\ell\rangle_{p_{tZ}\to 0}  \sim \Lambda
\left(\frac{M}{\Lambda}\right)^{\kappa
  \ln\frac{2+\kappa}{1+\kappa}}\,,\quad \kappa = \frac{2C_F}{\pi\beta_0}\,,
\end{equation}
where $\Lambda$ is the scale of the Landau pole in QCD, $M$ is the
invariant mass of the Drell-Yan pair,
$\beta_0 = (11 C_A - 2n_f)/(12\pi)$ and $C_F = 4/3$, $C_A=3$, while
$n_f$ is the number of light quark flavours.
With $n_f = 5$, this gives $\Lambda^{0.51} M^{0.49}$.
In practice this simple scaling is accurate only for large values of
$M$, and the result from a full NNLL resummation (green curve) can
be read off as the intercept of the corresponding curve in Fig.~\ref{fig:avg-jet-pt},
i.e.\ $2.5\GeV$, which coincides well with the intercept of the
simulations without MPI (blue curves).

Next consider the red curves in Fig.~\ref{fig:avg-jet-pt}, those with
MPI.
For high $p_{tZ}$ values, the leading jet $p_t$ again tracks
$p_{tZ}$.
However for low $p_{tZ}$ values, the average leading jet $p_t$
saturates at a value of about $10\GeV$, which is significantly above
the MPI-off result.
The interpretation is that in events with MPI, for low $p_{tZ}$, the
leading jet almost always comes from an MPI scatter, not from the hard
scatter, and it has a characteristic scale of the order of $10\GeV$.
This may be surprising if one thinks of MPI as genuinely
non-perturbative, but less so if one considers that Pythia
simulates MPI as semi-hard
scatterings~\cite{Sjostrand:1987su,Sjostrand:2004pf}.
(We found similar results with Herwig~7.2's
\cite{Bellm:2019zci,Bewick:2019rbu} implementation of a comparable
model~\cite{Bahr:2008dy,Bellm:2019icn}).

Fig.~\ref{fig:avg-jet-pt} provides the foundation for the rest of this
letter.
Specifically, if we consider events with a stringent cut on $p_{tZ}$,
we ensure the near total absence of hadronic radiation from the
primary scatter (defined as that producing the $Z$).
Existing experimental work confirms that the relative MPI contribution
is enhanced by choosing a low $p_{tZ}$ cut, for example using
$p_{tZ}< 5$ or $10\GeV$~\cite{CMS:2022vkb,CMS:2017ngy,Alioli:2016wqt,Bansal:2016iri,ATLAS:2014yqy}.
From Fig.~\ref{fig:avg-jet-pt} we observe that if we choose a $p_{tZ}$
cut that corresponds to the onset of the low $p_{tZ}$ plateau of the
MPI-off curves, i.e. $p_{tZ} < C_Z = 2\GeV$, we will obtain a near
optimal selection for the study of MPI:
if one takes $C_Z$ any higher, one increases contamination from
hadronic activity due to the primary hard scatter; if one takes it any
lower, there is no further advantage in terms of reducing primary
hard-scatter contamination, but one loses cross section (and also reaches
the limit of experimental lepton resolution).
Our choice selects about $4{-}5\%$ of the Drell-Yan events that pass the muon
cuts, i.e.\ a cross section after the $C_Z$ cut of about $40\pb$ at
$\sqrt{s}=13.6\TeV$.
For an LHC Run~3 luminosity of $300\fb^{-1}$, this would yield a
sample of about 12 million events.

At first sight, Fig.~\ref{fig:avg-jet-pt} might suggest that MPI
dynamics can be observed only at relatively low
$p_{tj}^\ell \sim 10\GeV$.
However after applying the $p_{tZ}$ cut, we can consider a much wider
array of observables, some of which extend over a range of jet
transverse momenta.
The simplest is the cumulative inclusive jet spectrum, i.e.\ the
average number of jets above some $p_{tj,\min}$, as a function of
$p_{tj,\min}$,
\begin{equation}
  \label{eq:2}
  \langle n(p_{tj,\min}) \rangle_{C_Z} =
  \frac{1}{\sigma(p_{tZ} < C_Z)} \int_{p_{tj,\min}}
  \!\!\! dp_{tj}
  \frac{d\sigma_\text{jet}(p_{tZ} < C_Z)}{dp_{tj}}\,.
\end{equation}
To a good approximation this observable is given by a straight sum of
the number of jets from the primary process and the number of jets
from the MPI.
The approximation is broken only by the potential overlap (in a cone
of size $R$ in rapidity and azimuth) of hadrons from the two scatters,
and the approximation is exact in the limit of small $R$.
Precisely for this reason, from here onwards we shall use $R=0.4$
rather than the $R=0.7$ of Fig.~\ref{fig:avg-jet-pt}.
All results ($R=0.4$ and $R=0.7$) use area
subtraction~\cite{Cacciari:2007fd,Cacciari:2008gn} to further reduce
the impact of such overlap, notably as concerns any underlying-event
pedestal of transverse momentum from the softest part of the MPI.
We include a jet rapidity cut, $|y_j|<2$, to mimic
the central acceptance of the ATLAS and CMS detectors.
\begin{figure}
  \centering
  \includegraphics[width=\columnwidth,page=1]{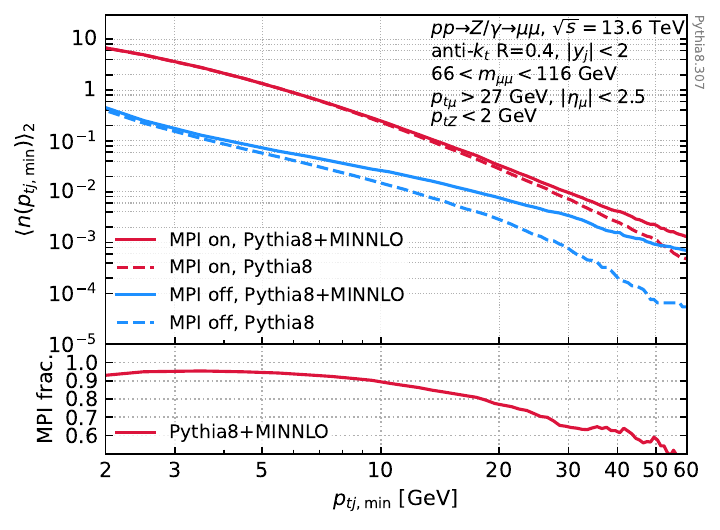}
  \caption{The cumulative inclusive jet spectrum
    $\langle n(p_{tj,\min}) \rangle_{C_Z}$ normalised to the number of
    events passing the cut $p_{tZ} < C_Z = 2\GeV$, with MPI on and off.
    The lower panel shows the fraction of jets that come from the MPI,
    demonstrating purity of $50{-}90\%$ across a broad range of
    $p_{tj,\min}$ jet cuts.  }
  \label{fig:cumul-incl}
\end{figure}

The cumulative inclusive jet spectrum is shown in
Fig.~\ref{fig:cumul-incl}.
It is clear that the vast majority of jets come from the MPI scatters
rather than the primary scatter, even for the relatively large value
of $p_{tj,\min} = 50\GeV$. 
Looking instead at moderately low $p_{tj,\min}$ values,
Fig.~\ref{fig:cumul-incl} indicates that on average there is one jet
with $p_t \gtrsim 6 \GeV$, which is
broadly consistent with the plateau at $10\GeV$ in
Fig.~\ref{fig:avg-jet-pt}, considering that Fig.~\ref{fig:cumul-incl}
uses $R=0.4$ instead of $R=0.7$, and that it has a limited
rapidity acceptance.
Note that for large $p_{tj,\min}$, the sample without MPI is dominated
by events where the $Z$ is accompanied by two opposing jets.
The Pythia8+MiNNLO sample includes the matrix element for that process
at leading order (LO), while Pythia8 does not, thus explaining the observed
difference between the two curves for $p_t \gtrsim 10\GeV$.

It is useful to define the pure MPI contribution to the cumulative
inclusive jet spectrum,
\begin{equation}
  \label{eq:n-pure}
  \langle n(p_{tj,\min}) \rangle_{C_Z}^\text{pure-MPI} \equiv
  \langle n(p_{tj,\min}) \rangle_{C_Z}
  -
  \langle n(p_{tj,\min}) \rangle_{C_Z}^\text{no-MPI}\,.
\end{equation}
In an actual experimental analysis, one might want to use a
next-to-leading order (NLO) $Z+2\text{jet}$ sample to subtract the
hard-event contribution.
Let us now see how Eq.~(\ref{eq:n-pure}) connects with the widely used
``pocket formula'' for double-parton scattering.
That formula states that the double-parton scattering cross section 
for two hard processes $A$ and $B$ is given by
\begin{equation}
  \label{eq:pocket}
  \sigma_{AB} = \frac{\sigma_A \sigma_B}{\sigmaeff}\,,
\end{equation}
where $\sigmaeff$ for $pp$ collisions is measured to be of the order of
$15{-}20\mb$~\cite{CDF:1997lmq,D0:2009apj,ATLAS:2013aph,CMS:2013huw,ATLAS:2018zbr,CMS:2021lxi,CMS:2022pio}  
(for processes involving a vector boson) and is
related to an effective area over which interacting partons are
distributed in the proton.
We take process $A$ to be $Z$ production with $p_{tZ} < C_Z$ and
process $B$ to be inclusive jet production, and consider a
$p_{tj,\min}$ that is large enough for the pocket-formula to be valid,
i.e.\ such that $\sigma_B/\sigma_\text{eff} \ll 1$.
This yields
(\cite{supplement}, \S~\ref{app:pocket})
\begin{equation}
  \label{eq:n-pure-pocket}
  \langle n(p_{tj,\min}) \rangle_{C_Z}^\text{pure-MPI}
  \simeq
  \frac{1}{\sigma_{\text{eff}}}
    \int_{p_{tj,\min}}
  \!\!\! dp_{tj}
  \frac{d\sigma_\text{jet}}{dp_{tj}}\,,
\end{equation}
where $\frac{d\sigma_\text{jet}}{dp_{tj}}$ is the inclusive jet cross
section for jet production, without any requirement that a $Z$ be
present in the event.\footnote{Using the Pythia minimum-bias process
  to generate the reference jet sample, we find
  $\sigma_\text{eff} \simeq 30 \mb$, somewhat larger than in standard
  measurements.
  \logbook{}{see eyeball fit on p.4 of ../../gavin-studies/13.6TeV-2023-03/inclusive-spect.pdf}
  This may imply that Pythia is underestimating the MPI or
  overestimating the minimum-bias jet spectrum,
  or that the data used for standard $\sigma_\text{eff}$ extractions
  has a higher level of MPI activity than would be seen with a
  $p_{tZ} < 2\GeV$ cut.  }
The right-hand-side of Eq.~(\ref{eq:n-pure-pocket}) does not involve
$C_Z$, and thus the pocket-formula prediction is that
$\langle n(p_{tj,\min}) \rangle_{C_Z}^\text{pure-MPI}$ should be
independent of $C_Z$.

The pocket formula is, however, known to be an approximation.
The difficulty of obtaining a pure MPI sample has so far limited the scope
for investigating more sophisticated theoretical predictions.
One particularly interesting effect not captured in the pocket formula
relates to perturbative interconnection between the primary scattering and the
secondary scattering, as in Fig.~\ref{fig:mixing-hard-proc}b, where at
least some of the partons entering the two separate hard scattering
processes ($Z$ and dijet production) have a common origin, e.g.\ a
perturbative $g \to q\bar q$ splitting, with the $\bar q$ involved in
$Z$ production and the $q$ involved in di-jet production.

Our procedure of constraining the $Z$ transverse momentum means that
the partons that annihilate to produce the $Z$ will almost always have
a low transverse momentum, which reduces the likelihood of their
having been produced in a perturbative splitting.
In contrast, if we relax the constraint on $p_{tZ}$, we will allow for
substantially more initial-state radiation  from the partons that go on
to produce the $Z$.
The ISR partons can then take part in a separate hard scatter, i.e.\
increasing the interconnection contribution to 2HS,
Fig.~\ref{fig:mixing-hard-proc}b.

To evaluate potential sensitivity to this effect, we examine the ratio
between the 2HS rate with loose ($C_Z = 15\GeV$) and tight
($C_Z = 2\GeV$) constraints on $p_{tZ}$,
\begin{equation}
  \label{eq:DGS-ratio}
  r_{15/2} = \frac{
    \langle n(p_{tj,\min}) \rangle_{15}^\text{pure-MPI}
  }{
    \langle n(p_{tj,\min}) \rangle_{2}^\text{pure-MPI}
  }
  \,.
\end{equation}
In each case the 2HS rate is normalised to the number of $Z$ bosons
that pass the selection cut.
With the pocket formula the ratio should be $1$, and so an
experimental measurement of $r_{15/2}$ has the potential to provide powerful
constraints on deviations from the pocket formula.
Note that with $C_Z = 15\GeV$, the pure-MPI jet fraction is predicted
by Pythia8+MiNNLO to be about $25\%$ at $p_{tj,\min}=40\GeV$
(\cite{supplement}, \S~\ref{app:r15-2-dShower}), which should be
adequate for a quantitative extraction of $r_{15/2}$.

\begin{figure}
  \centering
  \includegraphics[width=\columnwidth,page=1]{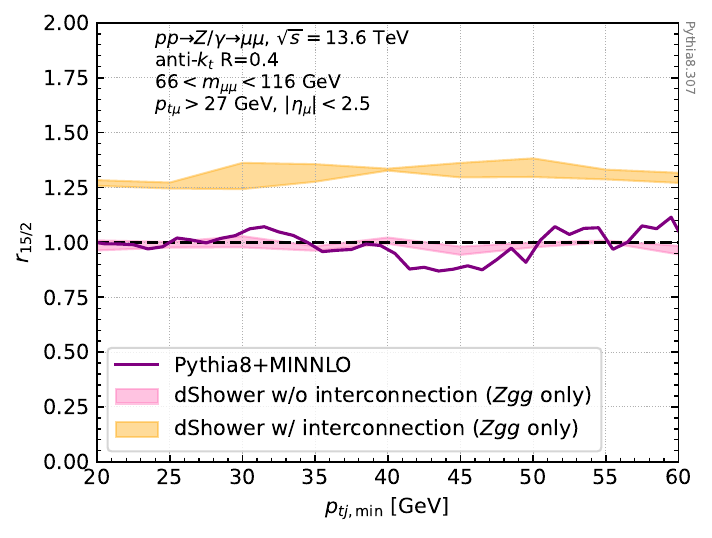}
  \caption{Pythia8+MiNNLO and dShower results for the $r_{15/2}$ ratio of
    Eq.~(\ref{eq:DGS-ratio}).
    Note the deviation from $1$  
    when perturbative interconnection is turned on between
    the primary and secondary hard scatters, i.e.\ diagrams as in
    Fig.~\ref{fig:mixing-hard-proc}b. 
    The dShower bands correspond to scale
    variation (see \cite{supplement} \S\ref{app:r15-2-dShower} for
    further details).
    They include  only the $Zgg$ final state, which
    represents about $50\%$ of independent 2HS, and so should be taken as
    qualitative.
    No jet rapidity cut is applied. 
  }
  \label{fig:dShower-results}
\end{figure}

Fig.~\ref{fig:dShower-results} shows the $r_{15/2}$ ratio evaluated in
three ways.
The Pythia8+MiNNLO curve corresponds to a full analysis,
using Pythia8+MiNNLO curve itself (without MPI), to evaluate the
no-MPI contribution for Eq.~(\ref{eq:n-pure}). 
Pythia8 does not include a perturbative interconnection mechanism
(though it has correlations related to momentum conservation and
colour reconnections~\cite{Christiansen:2015yqa}), and one sees a
result consistent with $r_{15/2} = 1$ to within statistical
fluctuations.

Fig.~\ref{fig:dShower-results} also includes curves from the dShower
program~\cite{Cabouat:2019gtm,Cabouat:2020ssr}.
This is a state-of-the-art code that simulates a pure 2HS component,
with the option of including interconnection effects according to
Ref.~\cite{Diehl:2017kgu}.
Rather than carrying out a full analysis (which would require a
consistent merging with a 1HS component), we determine the $r_{15/2}$
ratio based on the truth Monte Carlo information about the transverse
momentum of the hard outgoing partons in the $2\to2$ interaction, i.e.\
the second hard scattering.
The pink curve is the result without interconnection (with MSTW2008
PDFs~\cite{Martin:2010db}), and is 
consistent with $1$.
The orange curve includes interconnection effects, and one clearly
sees a 25-30\% violation of the pocket formula.
The scope for measuring this experimentally in a full analysis depends
critically on the systematic errors associated with the subtraction of
the no-MPI contribution in Eq.~(\ref{eq:n-pure}).
The significance of such a signal is discussed in \cite{supplement}
\S\ref{app:r15-2-dShower}, for various scenarios of uncertainties on
the no-MPI term, and the conclusion is that reasonable assumptions
lead to at least $2$ standard deviations at low $p_{tj,\min}$, which
would correspond to exclusion of the pocket formula.
The significance can be raised by increasing the accuracy of the
no-MPI predictions, e.g.\ with improved higher-order calculations.


The final question that we turn to is the sensitivity to more than two
simultaneous perturbative scatterings.
So far the only attempt to study this experimentally has been in
triple charmonium production, where the measured cross section has a
large uncertainty~\cite{CMS:2021qsn,Gaunt:2023kfk} and where generic difficulties in
understanding charmonium production complicate the interpretation of
the results.

Here we propose the study of charged-track jets, with moderately low
$p_t$ cuts.
To illustrate the study, we construct charged-track jets using 
charged particles with $|\eta| < 2.4$ and $p_t > 0.5\GeV$.
The use of charged particles enables the study of moderately low $p_t$
jets even in high-pileup runs, thus exploiting the full luminosity of
the LHC.
We order the jets in decreasing $p_t$, and first study the two leading
jets, with a ``product'' cut \cite{Salam:2021tbm},
$\sqrt{p_{t1} p_{t2}} > 9 f_{\text{chg}} \GeV$, and a ratio cut,
$p_{t2} > 0.6 \,p_{t1}$.
We quote the cuts in terms of a charged-to-neutral conversion ratio
$f_{\text{chg}} = 0.65$.
The overall scale of the cuts ensures a non-negligible likelihood that
each event contains at least one pair of jets.

\begin{figure}
  \centering
  \includegraphics[width=0.9\columnwidth,page=1]{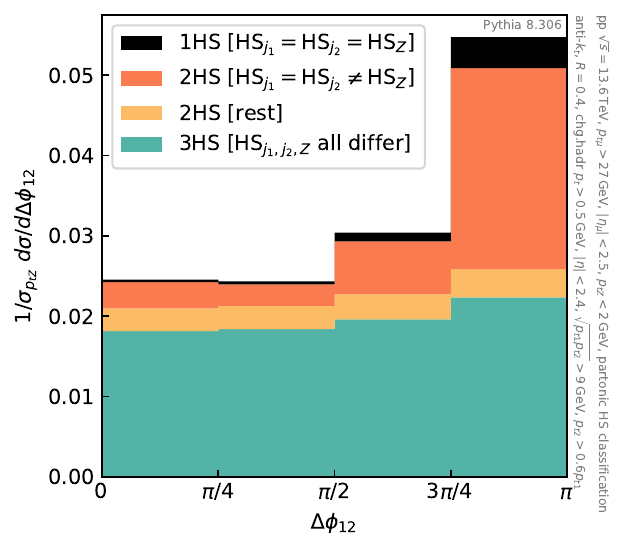}
  \caption{The distribution of the absolute value of
    $\Delta \phi_{12}$ between the two leading charged-track jets in
    events with $p_{tZ} < 2\GeV$ (cf.\ text for jet cuts).
    The plot shows a clear signal not just of 2HS (in the peak) but
    also of 3HS (plateau). }
  \label{fig:dijet-results}
\end{figure}

Fig.~\ref{fig:dijet-results} shows results for the absolute difference in
azimuthal angles between the two jets, $\Delta\phi_{12}$.
This observable is expected to peak around $\Delta\phi_{12} = \pi$
when the two jets come from the same hard partonic interaction, and to be
uniformly distributed between $0$ and $\pi$ when the two jets come
from distinct partonic interactions.
The plot clearly shows both a peak and a continuum component.
A parton-level based decomposition (\cite{supplement}
\S~\ref{sec:hard-scatter-perm}) of each histogram bin shows that the
plateau is dominated by events with 3 hard scatterings (3HS), where
each of the two leading jets comes from a different HS (each distinct
from the one that produced the $Z$).
The enhancement near $\Delta\phi_{12}=\pi$ originates mostly from 2HS
where the two jets are from a single HS that is distinct from the one
that produced the $Z$,  cf.\ Fig.~\ref{fig:mixing-hard-proc}a.
A measurement of $\Delta\phi_{12}$ would therefore provide clear and
quantifiable indications not only of 2HS but also of 3HS.

With such an unambiguous signal of 3HS, one may wonder if
it is possible to gain even further insight.
One obvious question is whether one can identify a system with two
back-to-back jets from one hard interaction and two further
back-to-back jets from another hard interaction, all distinct from the
$Z$ hard interaction, cf.\ Fig.~\ref{fig:mixing-hard-proc}c. 
This appears to be on the edge of feasibility, but also brings
sensitivity to 4HS (\cite{supplement}, \S~\ref{app:Z4j}).

To conclude, the study of Drell-Yan events with a tight cut on $p_{tZ}$ opens
the door to numerous new studies of multi-parton interactions, with
high-purity 2HS samples, sensitivity to the perturbative quantum field
theory effects that interconnect primary and secondary scatters, and
the scope for extensive investigations into 3HS and perhaps even beyond.
The studies outlined here are all possible with existing and Run~3
data.
The subset of studies that extends to relatively low jet $p_t$ values
should be feasible with charged-track jets. 
There is also ample scope for further exploration, for example in
terms of the choices of jet cuts, or studies in other collision
systems such as $p\,\text{Pb}$.
We expect experimental results on these questions to have the
potential for a significant impact not just on our intrinsic
understanding of multi-parton interactions but also for the accurate
modelling of hadron collisions that will be needed for the broad range of
high precision physics that will be carried out at the high-luminosity
upgrade of the LHC and at potential future hadron \mbox{colliders}.

\begin{acknowledgments}
  We are grateful to Jonathan Gaunt for extensive discussions and
  access to the dShower code, and to Lucian Harland-Lang and Paolo
  Torrielli for discussions in the early stages of this work.
  The work of JRA is supported by the STFC under grant ST/P001246/1.
  The work of PM has
  been funded by the European Union (ERC, grant agreement
  No. 101044599, JANUS).
  The work of LR is supported by the Swiss National Science Foundation
  contract PZ00P2$\_$201878.
  The work of GPS and ASO has been funded by the European Research
  Council (ERC) under the European Union's Horizon 2020 research and
  innovation program (grant agreement No 788223),
  and that of GPS additionally by a Royal Society Research
  Professorship (RP$\backslash$R1$\backslash$180112) and by the
  Science and Technology Facilities Council (STFC) under grant
  ST/T000864/1.
  This research was supported by the Munich Institute for Astro-, Particle and BioPhysics (MIAPbP)
  which is funded by the Deutsche Forschungsgemeinschaft (DFG, German Research Foundation) 
  under Germany's Excellence Strategy -- EXC-2094 -- 390783311.
  Views and opinions expressed are
  those of the authors only and do not necessarily reflect those of
  the European Union or the European Research Council Executive
  Agency. Neither the European Union nor the granting authority can be
  held responsible for them.
\end{acknowledgments}

\bibliographystyle{apsrev4-2}
\bibliography{Zpt}

\input{supplementary_material}

\end{document}

%% file: supplementary_material.tex
\newpage

\onecolumngrid
\newpage
\appendix

\section{Supplementary material}
Here we provide additional material to help the reader
reproduce the results described in the letter.

\subsection{The average transverse momentum of the leading jet when
$p_{tZ} \to 0$}
\label{app:scaling}
In this appendix, we provide a derivation of
Eq.~\eqref{eq:ptj-lead}. Our starting point is the following
equation~\cite{Bizon:2017rah,Monni:2019yyr}\footnote{See e.g. section
  3.2 and Appendix C of Ref.~\cite{Bizon:2017rah}.}
\begin{align}\label{eq:master}
\frac{d\sigma}{d p_t} = \sigma_0 p_t \int d b\,b\,J_0(p_t b)&\int
\frac{d k}{k}\,e^{-R(k)}R'(k) J_0(bk)\times \exp\left\{-R'(k)\int_0^{k}\frac{dq}{q}(1-J_0(bq ))\right\}\,,
\end{align}
describing the small-$p_t$ limit of the Drell-Yan spectrum. We proceed to evaluate Eq.~\eqref{eq:master}
at leading-logarithmic (LL) accuracy 
(radiation strongly ordered both in angle and
transverse momentum). This is sufficient to extract the qualitative scaling
given in Eq.~\eqref{eq:ptj-lead}. A more refined calculation with NNLL
accuracy can be performed using the double-differential resummation of
Ref.~\cite{Monni:2019yyr}, as shown in Fig.~\ref{fig:avg-jet-pt}. 

\begin{figure}
  \centering
   \includegraphics[width=0.45\textwidth]{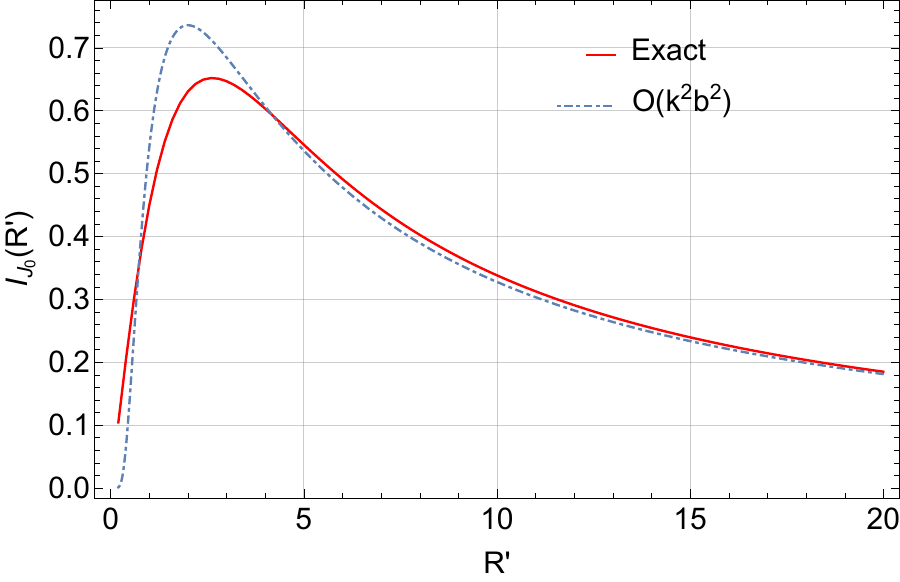}
   \includegraphics[width=0.45\textwidth]{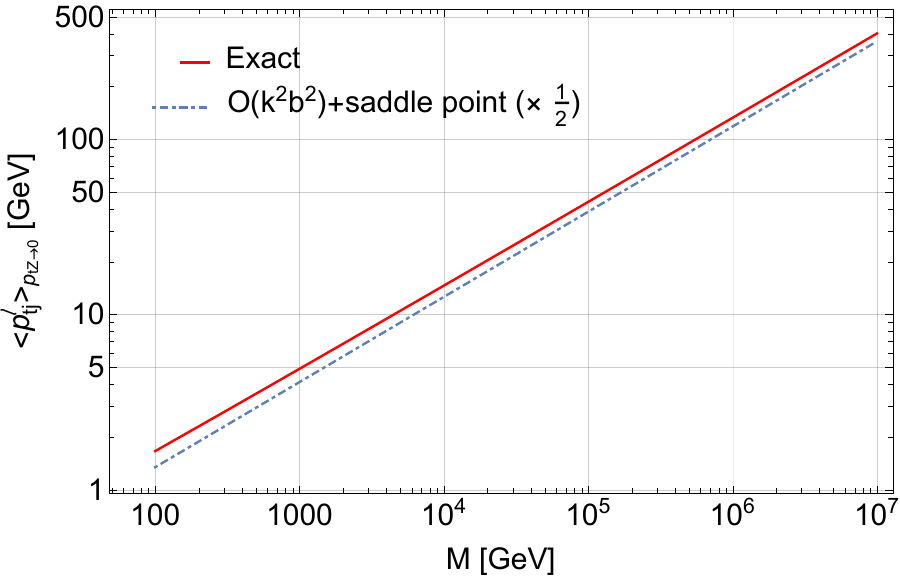}
  \caption{Left: comparison of the exact solution of the integral in
    Eq.~\eqref{eq:besselint2} to its analytic approximation in the
    r.h.s. of the same equation. Right: comparison of the exact
    numerical scaling for $\langle p_{tj}^\ell\rangle_{p_{tZ}\to 0} $
    to the analytic asymptotic estimate given in
    Eq.~\eqref{eq:finalscaling}.}
     \label{fig:besselint}
\end{figure}

At LL, the variable $k$ in Eq.~\eqref{eq:master} represents the
transverse momentum of the leading emission, which coincides with the
leading jet at this logarithmic accuracy. The quantity $R(k)$ is the
Sudakov radiator (see e.g. Ref.~\cite{Bizon:2017rah}) and
$R'(k)\equiv d R(k)/d \ln(M/k)$ with $M$ being the invariant mass of
the Drell-Yan pair. At LL it reads
\begin{equation}
R(k) = - \frac{2 C_F}{2\pi\alpha_s \beta_0^2} \left(2 \lambda +
  \ln(1-2\lambda)\right)\,,\quad \lambda = \alpha_s \beta_0 \ln\frac{M}{k}\,,
\end{equation}
with
\begin{equation}
  \alpha_s\equiv \alpha_s(M) =\frac{1}{2\beta_0\ln\frac{M}{\Lambda}}\,,
\end{equation}
being the one-loop coupling constant evaluated at $M$ with
$\beta_0 = (11 C_A - 2\,n_f)/(12\pi)$. For the sake of simplicity we
have neglected the effect of parton distribution functions, which are
encoded in the Born cross section $\sigma_0$. To proceed, we introduce
the asymptotic moments of the leading jet $p_t$ as follows
\begin{equation}
\langle\left(p_{tj}^\ell\right)^n\rangle_{p_{tZ}\to 0}  \,\equiv\, \lim_{p_t\to 0}\frac{\frac{d\sigma^{(n)}}{d p_t} }{\frac{d\sigma^{(0)}}{d p_t} }\,,
\end{equation}
where
\begin{align}\label{eq:master-two}
\frac{d\sigma^{(n)}}{d p_t} = \sigma_0 p_t \int d b\,b\,J_0(p_t b)&\int
\frac{d k}{k}\,k^n\,e^{-R(k)}R'(k) J_0(bk)\times \exp\left\{-R'(k)\int_0^{k}\frac{dq}{q}(1-J_0(bq ))\right\}\,.
\end{align}
To evaluate the quantity $\langle p_{tj}^\ell\rangle_{p_{tZ}\to 0} $
we start by taking the limit $p_{t}\to 0$ of the Bessel function
$J_0(p_t b)\to 1$. We then focus on the integral in the exponent which
reads
\begin{equation}\label{eq:besselint}
  \int_0^{k}\frac{dq}{q}(1-J_0(bq )) = \frac{b^2 k^2}{8}
  {}_2F_3\left(1,1;2,2,2;-\frac{b^2 k^2}{4}\right) = \frac{k^2
    b^2}{8} + {\cal O}(k^4 b^4)\,.
\end{equation}
In order to find an analytic estimate for
$\langle p_{tj}^\ell\rangle_{p_{tZ}\to 0} $, we can now expand the
hypergeometric function in Eq.~\eqref{eq:besselint} to second order
in powers of $b k$ as shown in the r.h.s. This reflects the fact
that the integral is dominated by finite and small $b k \sim 1$,
since the small $p_t$ limit is driven by radiation with non
vanishing transverse momentum (hence finite $b$ and small
$k$). Defining $y= b k$ we thus find
\begin{equation}\label{eq:besselint2}
I_{J_0}(R'(k)) \equiv \int d y\,y\, J_0(y)\,
\exp\left\{-R'(k) \frac{y^2}{8}
    {}_2F_3\left(1,1;2,2,2;-\frac{y^2}{4}\right)\right\}=
  4\frac{e^{-2/R'(k)}}{R'(k)} + {\cal O}\left(\frac{1}{R'(k)^2}\right)\,,
\end{equation}
where the approximation is obtained using the r.h.s. of
Eq.~\eqref{eq:besselint}.  We show the full numerical solution to the
l.h.s. of Eq.~\eqref{eq:besselint2} and the r.h.s. approximation in
Fig.~\ref{fig:besselint} (left plot). While the r.h.s. of
Eq.~\eqref{eq:besselint} is formally accurate for large $R'(k)$ we
observe that it provides a reasonable approximation also at moderately
low values of $R'(k)$, even for $R'(k)$ of order one, where one might
have had concerns about the neglected $1/(R')^2$ terms in
Eq.~\eqref{eq:besselint2}. 

We then evaluate the remaining integrals over $k$, that is
\begin{equation}
\int
\frac{d k}{k^3}\,k^n\,e^{-R(k)}R'(k) I_{J_0}(R'(k))\,,
\end{equation}
for $n=0,1$, using the saddle point method, which leads to
\begin{equation}\label{eq:finalscaling}
\langle p_{tj}^\ell\rangle_{p_{tZ}\to 0} \sim \Lambda
\left(\frac{M}{\Lambda}\right)^{\kappa
  \ln\frac{2+\kappa}{1+\kappa}}\,,\quad \kappa = \frac{2C_F}{\pi\beta_0}\,.
\end{equation}
Higher order corrections can be included in the above result, and
modify the asymptotic scaling at most by a normalisation factor and
subleading powers of $1/\ln\frac{M}{\Lambda}$. The asymptotic scaling
in Eq.~\eqref{eq:finalscaling} is compared to the full numerical
calculation of $\langle p_{tj}^\ell\rangle_{p_{tZ}\to 0} $ starting
from Eq.~\eqref{eq:master} as shown in Fig.~\ref{fig:besselint} (right
plot), which shows that the approximate solution captures the correct
slope as a function of $M$ in the asymptotically large $M$ limit.

\subsection{Average number of jets with MPI and pocket formula}
\label{app:pocket}
In this appendix we derive Eq.~\eqref{eq:n-pure-pocket}, which
describes the average number of jets arising from 2HS in the
pocket-formula approximation.
Our starting point is the pocket formula~\eqref{eq:pocket}, which, for
convenience, we recast as an explicit sum over the exclusive jet cross
sections of a given multiplicity in the two processes $A$ ($Z$
production with $p_{tZ} < C_Z$) and $B$ (inclusive jet production).
This reads
\begin{equation}
  \label{eq:pocket-excl}
  \sigma_{AB} = \sum_{i}\sum_{j}\frac{\sigma^{(i)}_A \sigma^{(j)}_B}{\sigmaeff}\,,
\end{equation}
where $i,j\geq 0$ and $\sigma^{(n)}_{A,B}$ denotes the cross section
for producing exactly $n$ jets in process $A,B$. The Drell-Yan cross
section with a cut $p_{tZ} < C_Z$ is then obtained as
\begin{align}
  \label{eq:totalDY}
    \sum_{i}\sigma_{A}^{(i)} =\sigma(p_{tZ} < C_Z)\,,
\end{align}
A few considerations about Eq.~\eqref{eq:pocket-excl} are in
order. 
The pocket formula is a sensible approximation in
situations where the total cross section for the production of one
or more jets in process $B$ is much smaller than the effective cross
section $\sigmaeff$. 
This is connected with the requirement that the total cross section
for process $A$ is preserved upon summing inclusively over the
jet multiplicities.  This translates to the following unitarity
condition
\begin{align}
  \label{eq:unitarity}
  \sum_{j=0} \sigma_{B}^{(j)}
  \simeq \sigma_{\text{eff}}\,,
\end{align}
where the value of $\sigma_{B}^{(0)}$ term, which must be positive, is
implicitly defined so as to ensure the equality.
We choose to use the symbol $\simeq$ in Eq.~\eqref{eq:unitarity}
instead of an equal sign so as to reflect the fact that the expression
is sensible only if the sum from $j=1$ upwards is much smaller than
$\sigma_\text{eff}$.

To obtain $\langle n(p_{tj,\min}) \rangle_{C_Z}^\text{pure-MPI}$ as
defined in Eq.~\eqref{eq:n-pure}, we then calculate the differential
inclusive jet spectrum from Eq.~\eqref{eq:pocket-excl}, and get
\begin{equation}
  \label{eq:pocket-spectrum}
  \frac{d \sigma_{AB}}{dp_{tj}} = \sum_{i}\sum_{j}\frac{1}{\sigmaeff}\left(\sigma^{(j)}_{B}\frac{d \sigma^{(i)}_{A}}{dp_{tj}}+\sigma^{(i)}_{A}\frac{d \sigma^{(j)}_{B}}{dp_{tj}}\right)\,,
\end{equation}
where ${d \sigma^{(n)}_{X}}/{dp_{tj}}$ is the differential inclusive jet 
spectrum in the subset of events of process $X$ with $n$ jets.
We now make use of Eqs.~\eqref{eq:totalDY},~\eqref{eq:unitarity} and obtain
\begin{equation}
  \label{eq:mpi-jets}
  \langle n(p_{tj,\min}) \rangle_{C_Z} =
  \frac{1}{\sigma(p_{tZ} < C_Z)} \int_{p_{tj,\min}}
  \!\!\! dp_{tj}
  \frac{d\sigma_{AB}}{dp_{tj}} \simeq \langle n(p_{tj,\min})
  \rangle_{C_Z}^\text{no-MPI} + \sum_{j}\frac{1}{\sigma_{\text{eff}}}
    \int_{p_{tj,\min}}
  \!\!\! dp_{tj}
  \frac{d\sigma^{(j)}_B}{dp_{tj}}\,,
\end{equation}
where the use of $\simeq$ in Eq.~(\ref{eq:mpi-jets}) follows from
Eq.~\eqref{eq:unitarity} and
\begin{equation}
 \langle n(p_{tj,\min})
  \rangle_{C_Z}^\text{no-MPI} 
  =
  \sum_{i}\frac{1}{\sigma(p_{tZ} < C_Z)} \int_{p_{tj,\min}}
  \!\!\! dp_{tj}
  \frac{d\sigma_A^{(i)}}{dp_{tj}}\,.
\end{equation}
Eq.~\eqref{eq:n-pure-pocket} then follows from Eq.~\eqref{eq:n-pure}.

\subsection{Notes on the extraction of  $r_{15/2}$}
\label{app:r15-2-dShower}
Here we further comment on the elements that enter into
Fig.~\ref{fig:dShower-results}.
In the Pythia8+MiNNLO case, we calculate
$\langle n(p_{t,\min}) \rangle_{C_Z}^\text{pure-MPI}$ defined in
Eq.~\eqref{eq:n-pure} by taking the difference between the simulation
with and without inclusion of the full tower of MPI.
For $p_{tj,\,\text{min}}\geq 20\GeV$, as considered in
Fig.~\ref{fig:dShower-results}, Pythia8 indicates that the dominant
contribution stems from events with at most two hard scatters.
\logbook{}{see gavin-studies/13.6TeV-2023-02
/deltaphi-plots-nMPI.pdf}
This observation justifies the comparison between the
Pythia8+MiNNLO result and a simulation of just double parton
scattering, as is provided by the dShower code.
We stress that dShower, 
unlike Pythia, simulates events with just a fixed number of scatterings, i.e. 
it provides only the pure 2HS component of the full MPI ladder. In particular, 
this means that no events with a single hard scattering are present in the
simulation.
The lack of this 1HS process strongly affects the relative
fraction of jets from the Z(+jets) process as compared to the
$2\to2$ process.
Therefore, instead of 
performing a full jet analysis, as done in Pythia8+MiNNLO, 
we extract $r_{15/2}$ by running dShower with different generation cuts ($p_{tj,\,\text{min}}$) 
on the jets produced by the second hard scattering ($pp\to jj$) 
accompanying the $pp\to Z$ process. We note that this procedure 
is not equivalent to the experimental definition of $r_{15/2}$ but allows us
to estimate the magnitude of interconnection effects with a state-of-the-art calculation. 

The dShower simulation relies on $3$ light active
 flavours. Specifically, for the pocket formula simulation we adopt
 the $n_f=3$ MSTW2008 set~\cite{Martin:2010db} and an effective cross
 section $\sigma_\text{eff}=18.5\,\mb$, while we use the $n_f=3$ DGS
 set~\cite{Diehl:2017kgu,Cabouat:2019gtm} to account also for the
  interconnection between the two scatterings.
Another point to note is that the dShower simulation includes only a
subset of partonic channels, specifically those contributing to a
$Zgg$ final state (thus the actual interconnected diagram differs from
that illustrated in Fig.~\ref{fig:mixing-hard-proc}b).
In independent 2HS, for the $p_{tj,\min}$ range shown in
Fig.~\ref{fig:dShower-results}, the $gg$ channel represents
$50{-}60\%$ of the total second-hard scatter dijet rate.
\logbook{}{run ./cross-section -dijets -ptmin_gen 60 -pythia8 -noISR
  -noFSR -parton -nev 1e3 -noUE -out a; find 52\% $gg \to gg$ with ptmin of 60,
  61\% with ptmin of 20}
We expect that other partonic
 channels would see a comparable degree of interconnection, resulting
 in a similar value for $r_{15/2}$ across all channels, however this
 point clearly deserves to be verified with an explicit
 calculation.
Thus the dShower code gives an indication of the order of magnitude of
perturbative interconnection effects, but does not, as yet, predict
the full quantitative picture.

We estimate the theoretical uncertainty on the dShower prediction by varying 
for each of the two processes: (i) in the case when 
interconnection effects are on, the cutoff on the impact parameter, i.e. the $\nu$-scale in 
Ref.~\cite{Cabouat:2019gtm}, and (ii) the renormalisation scale
($\mu_R$), which also 
affects the definition of the shower starting scale. 
More precisely, we set dShower's \texttt{ParamNu} parameter 
to be either 1 or 0.5, corresponding to an impact parameter cutoff of the order of the
hard scale or half of it, respectively. Regarding $\mu_R$,  we set the flag \texttt{MuRisPT} to be 
either \texttt{True} or \texttt{False}, corresponding to setting the renormalisation 
scale to be of the order of either the invariant mass of the two scatterings or the
transverse momentum scale of the outgoing particles (the leptons in the DY case). 
In all cases, we run with \texttt{UnequalScale=True} so that the shower starting scales in the
two hard scatterings are independent. The envelope of all these variations 
constitutes the uncertainty band displayed in Fig.~\ref{fig:dShower-results}.

\begin{figure}
  \centering
  \includegraphics[width=0.4\textwidth]{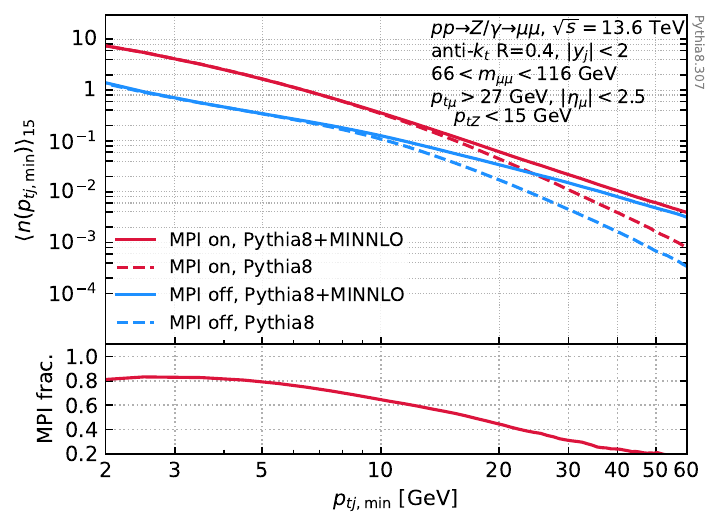}
  \caption{The analogue of Fig.~\ref{fig:cumul-incl} in the letter but
    for a $p_t$ cut on the $Z$ of $p_{tZ} < C_Z = 15\GeV$.
    This shows that at moderately high $p_t$, the MPI fraction is
    still reasonable, $\sim 25\%$. 
    As discussed in the text, 
    such an MPI fraction should still allow for a quantitatively
    reliable extraction of $\langle n(p_{t,\min}) \rangle_{15}^\text{pure-MPI}$, as needed
    for the evaluation of the $r_{15/2}$.
    The $Z$ selection results in a Pythia8+MiNNLO cross section of $\sigma_{p_{tZ}<
      15\GeV}\simeq 450 \pb$. 
  }
  \label{fig:nptmin-15}
\end{figure}

\begin{figure}
  \centering
   \includegraphics[width=0.4\textwidth]{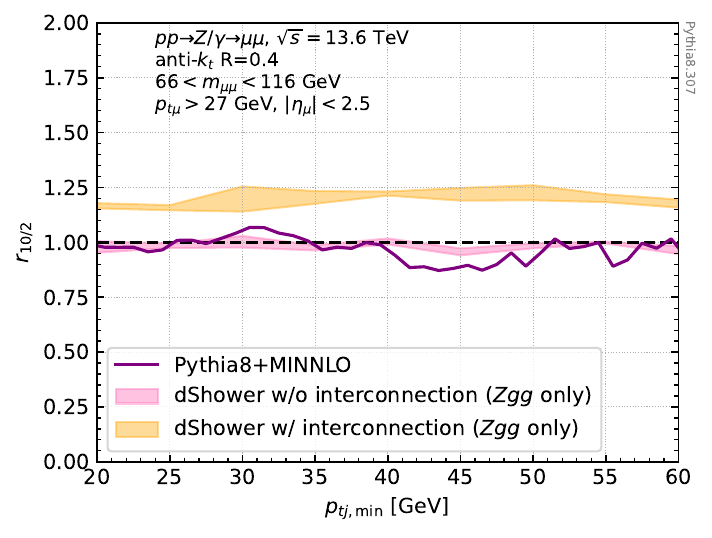}
  \includegraphics[width=0.4\textwidth]{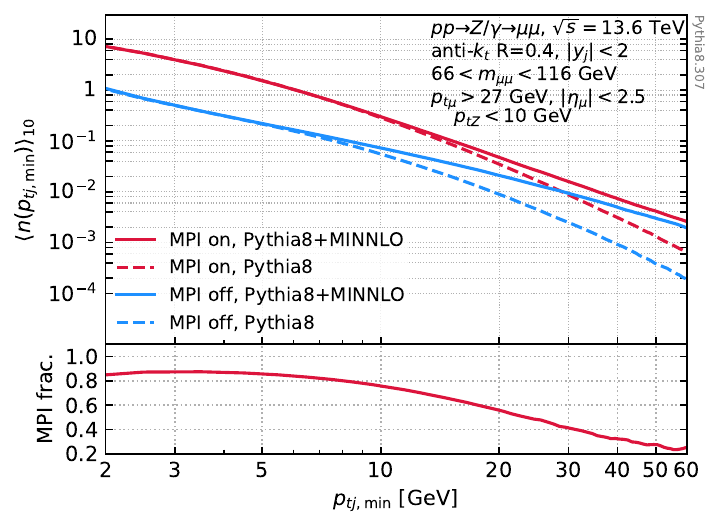}
  \caption{Same as Figs. \ref{fig:dShower-results} and \ref{fig:nptmin-15} but
    for a $p_t$ cut on the $Z$ of $p_{tZ} < C_Z = 10\GeV$.
    This shows higher MPI fraction but lower interconnection effects
    across all values of $p_{t,\min}$.
    The $Z$ selection results in a Pythia8+MiNNLO cross section of
    $\sigma_{p_{tZ}< 10\GeV}\simeq 340 \pb$.
    %
  }
     \label{fig:r10/2-nptmin-10}
\end{figure}

\begin{figure}
  \centering
   \includegraphics[width=0.4\textwidth]{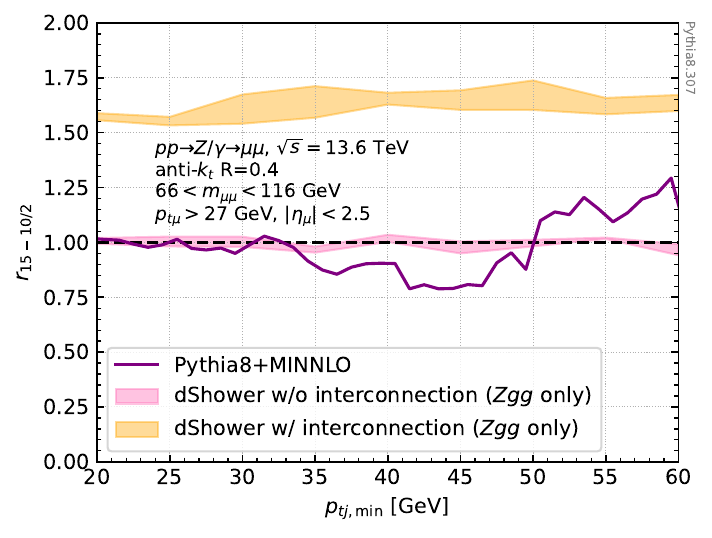}
  \includegraphics[width=0.4\textwidth]{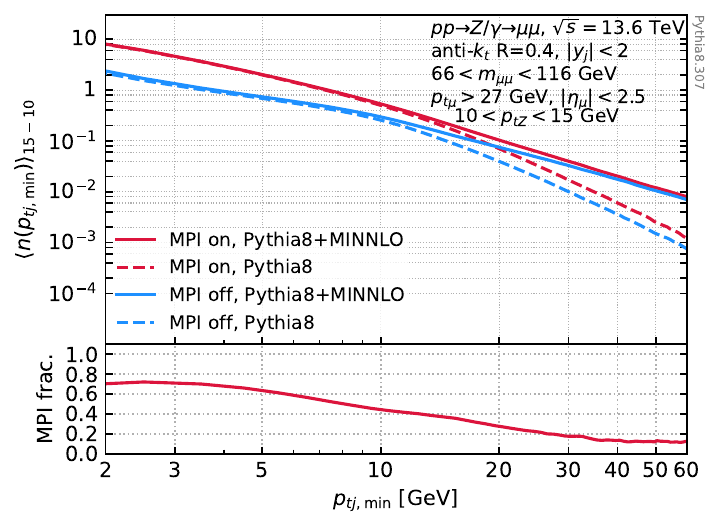}
  \caption{Same as Figs. \ref{fig:dShower-results} and \ref{fig:nptmin-15} but
    for a $p_t$ cut on the $Z$ of $10< p_{tZ} < 15\GeV$.
    This shows lower MPI fraction but higher interconnection effects
    across all values of $p_{t,\min}$.
    The $Z$ selection results in a Pythia8+MiNNLO cross section of
    $\sigma_{10<p_{tZ}<15\GeV} \simeq 110 \pb$. 
  }
     \label{fig:r1510/2-nptmin-1510}
\end{figure}

One general concern in the extraction of the $r_{15/2}$ ratio,
Eq.~(\ref{eq:DGS-ratio}), is whether the relative MPI contribution to
$\langle n(p_{t,\min}) \rangle_{15}$, shown in Fig.~\ref{fig:nptmin-15},
is sufficiently large that one can reliably determine $\langle n(p_{t,\min}) \rangle_{15}^\text{pure-MPI}$ 
after subtracting the no-MPI contribution, including its uncertainties.
In particular, one might wonder whether it would be beneficial to
lower the upper $p_{tZ}$ cut.
Fig.~\ref{fig:r10/2-nptmin-10} demonstrates the higher purities that
can be achieved by lowering the loose cut to $10 \GeV$, but at the
cost of a reduced impact of interconnection effects.
Alternatively, 
one can select a bin in $p_{tZ}$, e.g. $10<p_{tZ}<15 \GeV$ as shown in 
Fig.~\ref{fig:r1510/2-nptmin-1510}. In this case, we find an enhanced signal
of interconnection effects but low purities.
At these low purities, even a small offset in jet energies between
 the no-MPI and the MPI samples (e.g.\ due to imperfections of the
 area subtraction) may result in enhanced systematic errors on the
 $r_{x/2}$ determination.
This, combined with an enhanced sensitivity to statistical fluctuations,
may be the cause of the apparent deviation of the Pythia+MiNNLO
$r_{x/2}$ result from one at high $p_{tj,\min}$.

An analysis of the uncertainties can help us understand which choice
of cuts might give the most significant determination of deviations
from the pocket formula.
In particular one should examine how $r_{x/2}$ would be determined
experimentally, 
\begin{equation}\label{eq:rx2def}
	r_{x/2} = \frac{j_x^{\rm exp} - h^{\rm th}_x}{j_2^{\rm exp} - h^{\rm th}_2} \, ,
\end{equation}
with $j^{\rm exp}$ being the measured jet rate and  $h^{\rm th}$ the
theoretically determined no-MPI rate. The latter is subject to a theoretical 
uncertainty, which we write as $\Delta h^{\rm th} = f h^{\rm th}$, with $f$
being a fractional error. The measured jet rate is affected both by 
statistical and systematic uncertainties. We assume that the 
experimental systematic uncertainties would largely be correlated and 
so cancel in $r_{x/2}$. We have checked that the statistical uncertainty 
is much smaller than the theory uncertainty for an integrated luminosity of 
$300 \, \rm{fb}^{-1}$.~\footnote{The impact of the statistical uncertainty should be properly assessed 
in the case of a dedicated low-pileup run. }
Therefore, in what follows, we only consider the propagation of 
$\Delta h^{\rm th}$ into the $r_{x/2}$ uncertainty.

We find
\begin{equation}
\label{eq:delta-th-stat}
\Delta_{\rm th}^2 r_{x/2}  =
%
%
\left[f_2^2 \frac{(1-p_2)^2}{p_2^2}
  (r_{x/2})^2-2 f_2 f_x \rho \frac{(1-p_2) (1-p_x)}{p_x p_2}  r_{x/2} + f_x^2 
   \frac{(1-p_x)^2}{p_x^2}\right],
\end{equation}
where $\rho$ quantifies the correlation between the theory error at $C_Z=x\GeV$ and at $C_Z=2\GeV$. 
We remind the reader that the pocket-formula Eq.~\eqref{eq:n-pure-pocket}, corresponds to $r_{x/2}=1$.
In the following, we use as our estimate of $r_{x/2}$ the dShower result.
The theoretical uncertainty is expressed in terms of 
the MPI fraction or ``purity": $p=n^\text{pure-MPI}/(n^\text{pure-MPI}+h^{\rm
  th})$, as shown in the lower panels of Figs.~\ref{fig:cumul-incl}
and \ref{fig:nptmin-15}. (The purity is defined as that obtained when
$r_{x/2}=1$, and is always as obtained with a $|y_j| < 2$ cut). 

\begin{figure}
\centering
\includegraphics[width=0.3\textwidth, page=7]{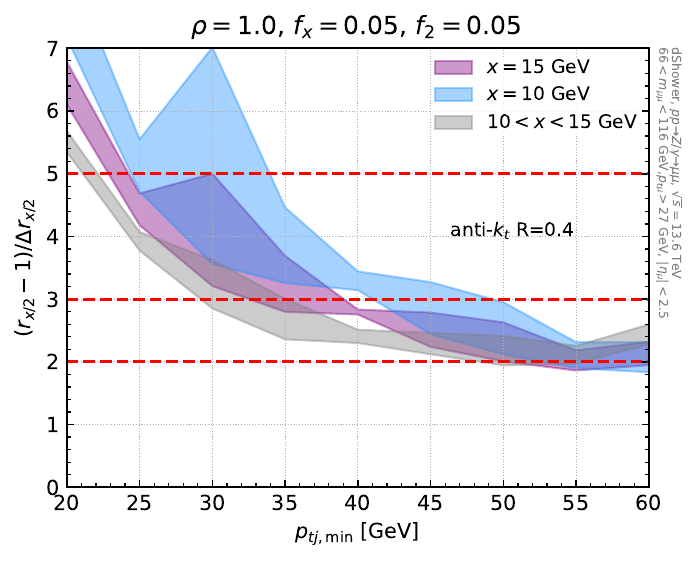}  \includegraphics[width=0.3\textwidth, page=6]{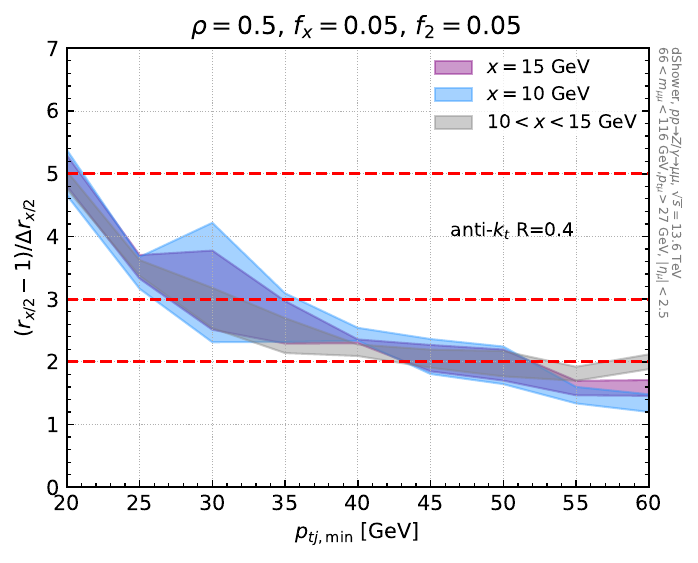} \includegraphics[width=0.3\textwidth, page=5]{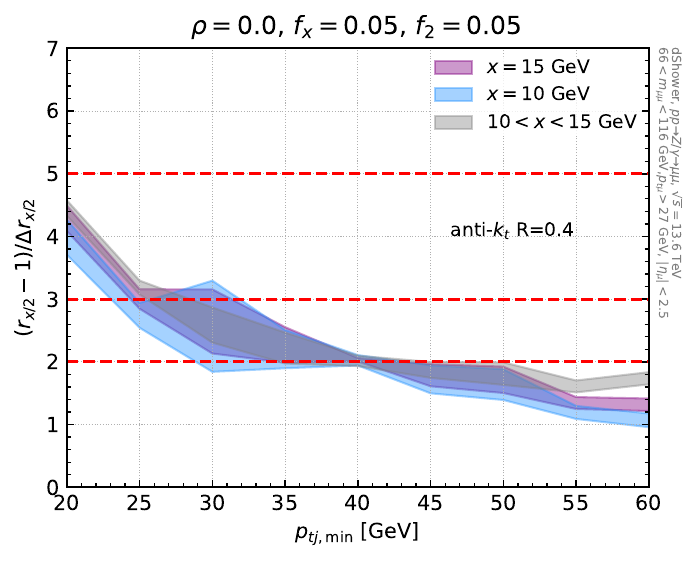} \\
\includegraphics[width=0.3\textwidth, page=7]{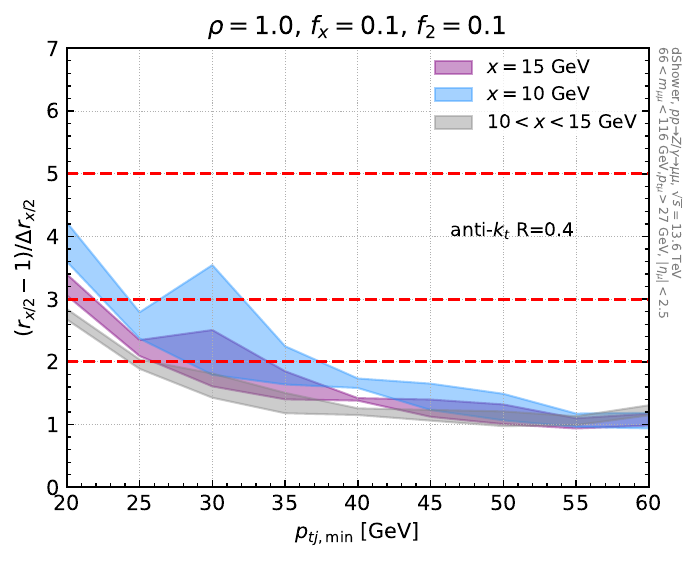}\includegraphics[width=0.3\textwidth, page=6]{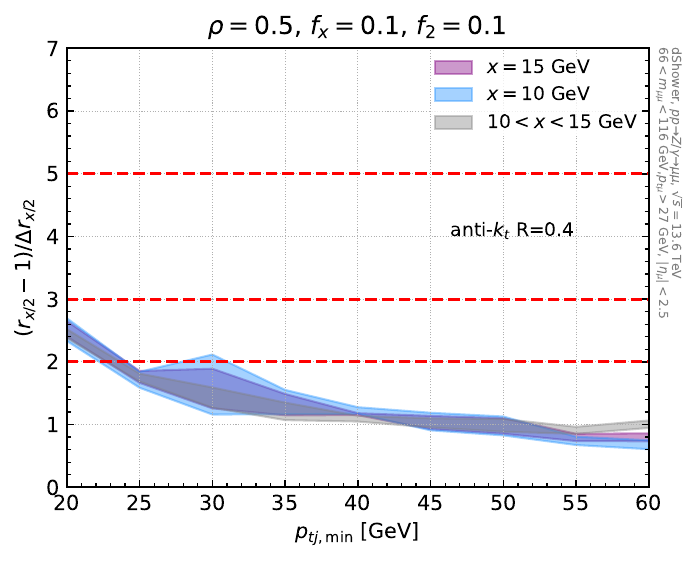} \includegraphics[width=0.3\textwidth, page=5]{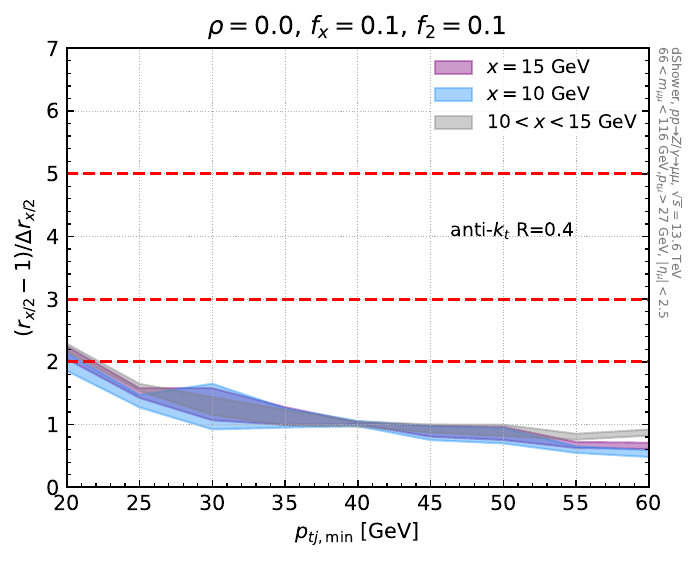} \\
\includegraphics[width=0.3\textwidth, page=7]{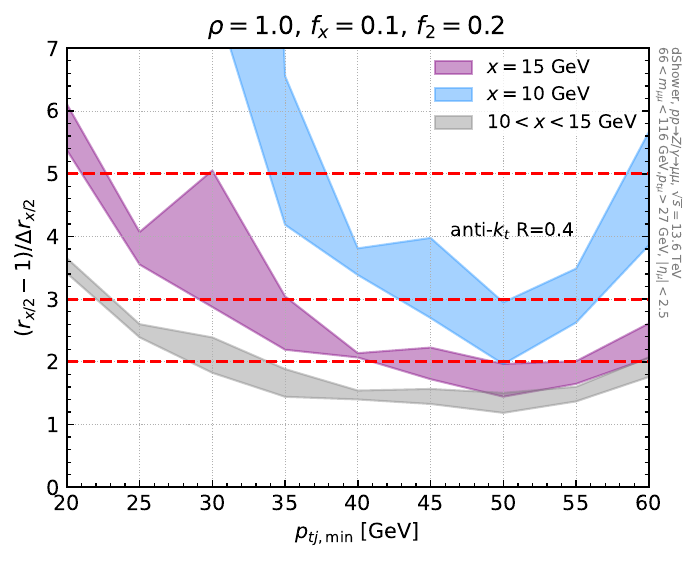}  \includegraphics[width=0.3\textwidth, page=6]{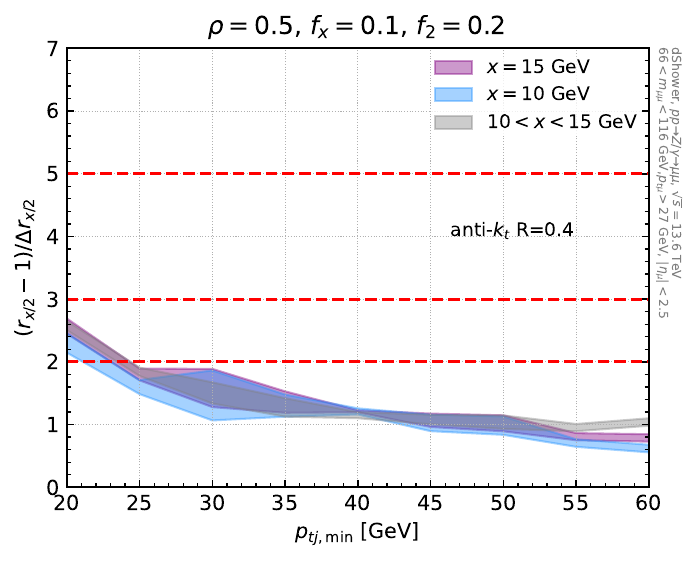} \includegraphics[width=0.3\textwidth, page=5]{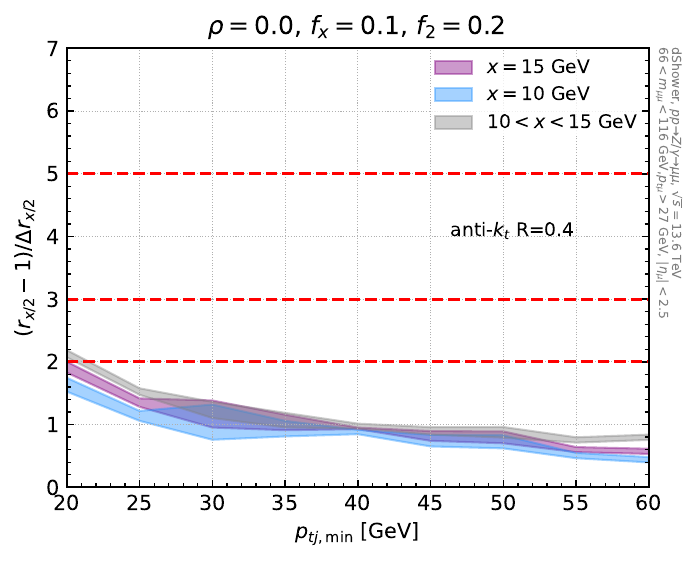} \\
\caption{Statistical significance of the detection of the breaking of
  the pocket-formula with the $r_{x/2}$ observable for three different
  values of $\rho$ in Eq.~\eqref{eq:delta-th-stat} (one per column)
  and three different assumptions for the fractional
    uncertainties, $f_x$ and $f_2$, on the no-MPI
    cross section (one per row).
  See main text for further
  details.
} 
 \label{fig:delta-r}
\end{figure}

The quantity of interest is the significance (number of $\sigma$)
for observing effects that go beyond the pocket-formula.
It can be obtained as $(r_{x/2}-1)/\Delta r_{x/2}$. 
We use a dShower-like signal as a baseline.
We plot the significance in Fig.~\ref{fig:delta-r}.
The three bands show distinct choices of the upper $p_{tZ}$ cut,
$15\GeV$ (as used in the main text), $10\GeV$ and $10<p_{tZ}<15\GeV$.
The columns show three values of the $\rho$ parameter when evaluating
the theoretical systematic uncertainty via Eq.~\eqref{eq:delta-th-stat}:
0 ($f_x$ and $f_2$ are fully uncorrelated), 0.5 and 1 (full
correlation).
The different rows show a range of assumptions about the theoretical
errors on the hard component:
the choices $f_x=f_2=5-10$\% mimic the expected accuracy of the NNLO
calculations of the $Z+2\text{-jet}$ rate that should hopefully become
available in the next few years.
The $f_x=10$\% and $f_2=20$\% reflects the uncertainty associated with
scale variations in the Pythia8+MiNNLO samples.
The observed significance of the deviation from the pocket formula
depends strongly both on the assumptions for $f_2$ and $f_x$ and on
the value of the correlation parameter $\rho$, as well as on
$p_{tj,\min}$.
In the optimistic $f_x=f_2=5\%$ scenario, at the lower end of the
$p_{tj,\min}$ range, we see at least $4\sigma$ significance even with
$\rho=0$, and over $5\sigma$ if some correlation is assumed.
Other scenarios still all give at least $2\sigma$ at low
$p_{tj,\min}$, which would be sufficient to exclude the pocket-formula
in the presence of an $r_{x/2}$ effect of the size suggested by
dShower.
The much larger significances in the lower left-hand plot are an
artefact of an almost exact cancellation of correlated uncertainties
between different $p_{tZ}$ cuts.
Overall, we see the choice of upper $p_{tZ}$ cut is not too critical.
A final comment is that the generally improved significance with small
$f_2$ and $f_x$ may provide an additional motivation for $Z+2$-jet
calculations at NNLO and beyond.

%

\subsection{$Z$ plus four-jet study}
\label{app:Z4j}

\begin{figure}
  \centering
  \includegraphics[width=0.45\columnwidth,page=2]{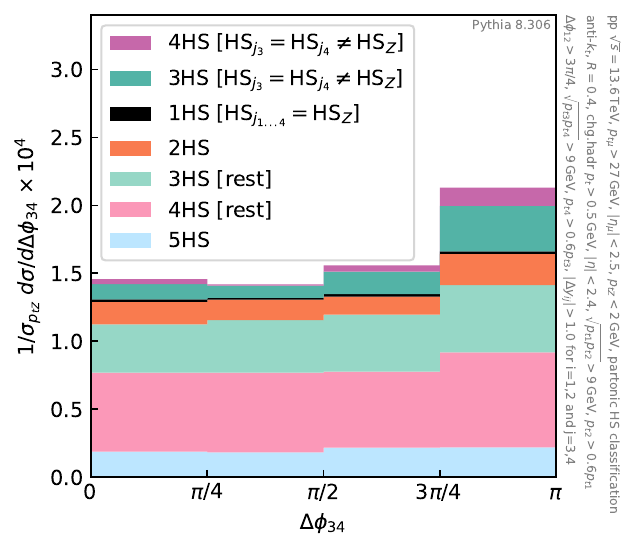}
  \caption{The distribution of $\Delta \phi_{34}$ from the four-jet
    study in the text, illustrating the rich decomposition into
    different numbers of MPI.}
  \label{fig:four-jet-results}
\end{figure}

Figure~\ref{fig:four-jet-results} shows a $Z$ plus four-jet study that
is intended to help examine the structure of 3HS, in particular
the situation where the $Z$ and each of the two pairs of jets arises
from distinct hard scatterings.
We apply the usual $p_{tZ} < C_Z = 2\GeV$ requirements,
and the same cuts for the two highest $p_t$ jets as in
Fig.~\ref{fig:dijet-results}, but with an additional constraint of
$\Delta \phi_{12} > 3\pi/4$, so as to enhance the contribution from
the situation where the two leading jets are from the same hard
interaction.
We then apply product and ratio cuts to a second pair of jets, jets
$3$ and $4$, $\sqrt{p_{t3} p_{t4}} > 9 f_{\text{chg}} \GeV$,
$p_{t4} > 0.6 \,p_{t3}$.
We also apply a rapidity cut $|\Delta y_{i,j}| > 1$, with $i=\{1,2\}$ and
$j=\{3,4\}$, to reduce the likelihood that a jet in the first pair and
a jet in the second pair originate from the fragmentation of a single
hard parton.
Finally, we plot the distribution of $\Delta \phi_{34}$ in
Fig.~\ref{fig:four-jet-results}.
We see some degree of peak around $\Delta \phi_{34} = \pi$, for the
most part a consequence of the 3HS that we were trying to isolate.
Meanwhile the plateau region receives contributions from a mix of 3HS,
4HS and even some 5HS, illustrating the considerable potential of such
a $Z+4$-jet analysis.
Clearly it would be interesting, in both the dijet and 4-jet studies,
to further investigate the structure of different numbers of
interactions, for example by varying the jet $p_t$ cuts so as to
modify the relative contributions from different numbers of hard
interactions.

\subsection{Determination of hard-scattering jet permutations}
\label{sec:hard-scatter-perm}

Several steps are required in order to obtain the breakdown into
numbers of hard scatterings shown in Figs.~\ref{fig:dijet-results} and
\ref{fig:four-jet-results}.

\begin{figure}
  \centering
  \includegraphics[width=\textwidth]{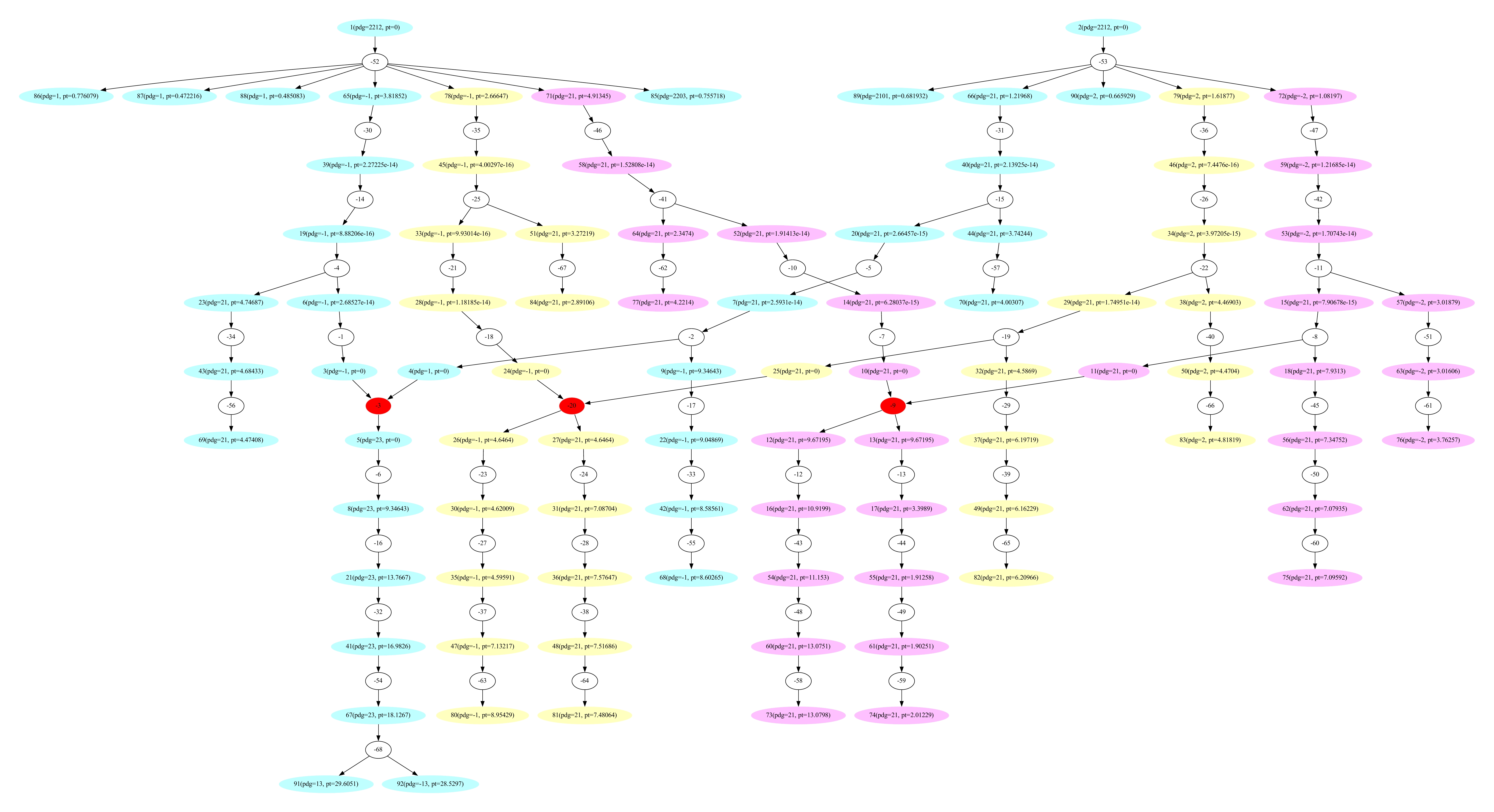}
  \caption{A graph representation of the branching and scattering in
    a parton-level event, as simulated with Pythia8.
    Each red region corresponds to a hard partonic scattering.
    The showering associated with production of the Drell-Yan pair is
    shown in cyan, while the showering associated with each of the two other 
    hard scatters (which are both $2\to2$ processes) is shown
    respectively in
    magenta and yellow.
    Typical Pythia8 events contain significantly more than two
    additional hard scatters, but the number has been restricted in
    this graph for simplicity, and simulation of final-state radiation
    has been turned off for the same reason.  }
  \label{fig:MPI-graph}
\end{figure}

In a parton-level Monte Carlo simulation with Pythia8, it is possible
to associate each parton with a specific underlying hard scattering.
To do so, we use the event record as represented through the HepMC2
package~\cite{Dobbs:2001ck}.
In identifying the hard scattering association of each parton, some
care is required, for example, to make sure that initial-state
radiation (and its subsequent showering) is correctly treated.
This is illustrated in Fig.~\ref{fig:MPI-graph}, which shows an event
with three hard scatterings (each represented in red).
All partons in the event (both intermediate and final) are
colour-coded according to their associated hard scatter.
One sees that final partons (those with no further vertices emanating from them)
may have their origin both before and after the hard scattering.

Given a hard scattering association for each parton, the next step is
to obtain a hard-scattering association for a given jet.
Our approach is, for each jet, to identify the fraction of the jet's
momentum that comes from each of the hard scatters.
We declare the hard scatter that contributes the most to the jet to be
the main source of that jet.

The next issue is that of how to transfer the information to
hadron-level analyses such as those in Figs.~\ref{fig:dijet-results}
and \ref{fig:four-jet-results}.
Ideally, one would want to be able to identify, for each hadron, which
MPI it came from.
However hadrons may come from more than one MPI, for example due to
colour reconnections~\cite{Christiansen:2015yqa}.
Therefore to obtain Figs.~\ref{fig:dijet-results} and
\ref{fig:four-jet-results}, we carry out two analyses: one at hadron
level, which determines the normalisation of each bin, and one at
parton level, which determines the relative contributions of different
hard-scattering permutations to each bin.

\begin{figure}
  \centering
  \includegraphics[width=0.4\textwidth]{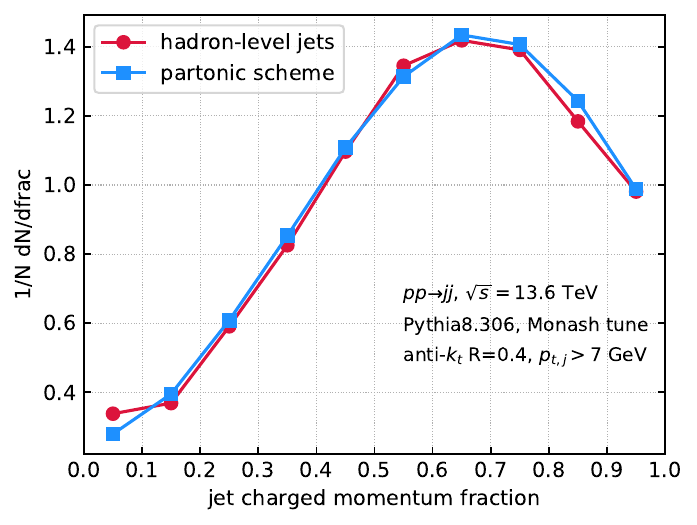}
  \caption{Illustration that the procedure described in the text to split partons into
    collinear pieces, some of which are taken
    charged (``partonic scheme''), correctly 
    reproduces the hadron-level distribution of charged-to-full 
    transverse momentum fraction in jets.
    The $p_{t,j}>7\GeV$ cut is applied on the full jets.
    See text for further details.
  }
  \label{fig:charged-games}
\end{figure}

If the hadron-level analysis uses all hadrons, we believe the above
procedure to be adequate.
However Figs.~\ref{fig:dijet-results} and \ref{fig:four-jet-results}
use only charged hadrons, which introduces extra fluctuations (for
example changing the $p_t$ ordering of the jets).
To reflect this in the parton-level analysis, we adopt a heuristic
approach that splits each parton collinearly into three or four pieces
(with equal probability), distributes the parton's momentum randomly
between the different pieces, and then assigns each piece a non-zero
charge with a $61\%$ probability (we do not impose charge
conservation).\footnote{Note that we take a slightly larger
  $f_\text{chg} = 0.65$ in the main text when translating full-jet
  cuts to charged-track jet cuts.
  This is to compensate in part for the fact that with a steeply
  falling spectrum, the cuts favour jets in which the charged
  component fluctuated up.}
To test the ability of such a procedure to correctly simulate
charged-to-full fluctuations, we take two samples of simulated jets,
one at hadron level, the other at parton level.
In each case we require the full jet to have a minimum $p_t$ of
$7\GeV$.
In the hadron-level sample, we examine the distribution of the ratio
of the charged-hadrons' total $p_t$ in each jet to the full jet $p_t$.
In the parton-level sample, we examine the distribution of the ratio
of the ``charged'' parton pieces' total $p_t$ in each jet to the full
jet $p_t$.
The two distributions are shown in Fig.~\ref{fig:charged-games} and
can be seen to be remarkably similar.
In determining the relative fractions of different hard-scattering
permutations for a given bin of Figs.~\ref{fig:dijet-results} and
\ref{fig:four-jet-results}, we use jets obtained from the clustering
of just the charged parton pieces.
We have verified that the histograms (summed over all numbers of hard
scatters) in the 2-jet and 4-jet analyses have similar shapes in the
charged-hadron and charged-parton-piece analyses.
We do, however, find a difference in overall normalisation, by a factor of
about $1.5{-}2.5$ which is expected, because full hadron-level jets
tend to have less energy than the full parton level jets, and the
splitting of partons into collinear pieces does not correct for
that.
We do not expect this to significantly modify the relative fractions
of different hard-scattering permutations in
Figs.~\ref{fig:dijet-results} and \ref{fig:four-jet-results}.
